\documentclass[prb, twocolumn, superscriptaddress, showpacs,floatfix,letterpaper]{revtex4-1}  
\usepackage{hyperref}
\usepackage{amssymb}
\usepackage{amsmath}
\usepackage{float}
\usepackage{bbm}
\usepackage{graphicx}
\usepackage{epsfig}
\usepackage{epstopdf}
\usepackage[usenames]{color}

\newcommand{\ez}{\hat{e}_z}

\begin{document}
\bibliographystyle{apsrev4-1}

\title{Superconducting and Pseudogap effects on the interplane conductivity and  Raman scattering cross section in the two dimensional Hubbard Model}

\author{E. Gull}
\affiliation{Department of Physics, University of Michigan, Ann Arbor, Michigan 48109, USA}

\author{A. J. Millis }
\affiliation{Department of Physics, Columbia University, New York, New  York 10027, USA}

\date{\today }

\begin{abstract}
Cluster dynamical mean field methods are used to calculate the superconductivity-induced changes in the  interplane conductivity and  Raman scattering cross section of the two dimensional Hubbard model. When superconductivity emerges from the pseudogap, the superconducting response is found to be diminished in amplitude, broadened and, in the case of the interplane conductivity, shifted to higher frequency.  The results are in  agreement with data on high temperature copper-oxide superconductors indicating that the Hubbard model contains the essential low energy physics of the pseudogap and its interplay with superconductivity in the cuprates.  \end{abstract}

\pacs{
74.25.nd,%Raman and optical spectroscopy (superconductivity)
74.72.Kf,%Cuprates/Pseudogap regime 
74.20.-z,%Theories and models of superconducting state
74.25.Dw,%Superconductivity phase diagrams
}

\maketitle
\section{Overview}
Three important characteristics of the layered copper oxide materials such as YBa$_2$Cu$_3$O$_{6+x}$ are the existence of a large-gap insulating phase when $x=0$, of a ``pseudogap'' regime involving a suppression of the density of states, for a range of $x>0$, and of a $d_{x^2-y^2}$-symmetry superconducting state occurring for a range of $x$ which partly overlaps the pseudogap regime. Understanding the relation (if any) between these phenomena is one of the central issues in the field. Two  experimental probes which have been important in the discussion  of this issue are the interplane (c-axis)  conductivity and the $B_{1g}$ Raman scattering cross section. These  spectroscopies are interesting because they appear to be controlled by the ``antinodal'' electrons which are most affected by superconductivity and the pseudogap and exhibit dramatic variations with temperature and carrier concentration. Both of these probes reveal striking temperature and $x$-dependent effects associated with the pseudogap and superconductivity, which one would like to understand theoretically.  The problem has attracted considerable attention, but  work to date has been primarily based on approximate analytical approaches, often involving particular phenomenologically chosen ans\"{a}tze for the relevant physical processes. 

The development of cluster dynamical mean field theory \cite{Hettler98,Hettler00,Lichtenstein00,Kotliar01,Maier06} has changed the theoretical situation, providing an unbiased (in the sense of not pre-selecting a particular interaction channel or set of diagrams) numerical approach to determining the properties of some of the basic models of condensed matter physics. Recent algorithmic developments \cite{Gull08,Gull10_submatrix,Gull11} enable implementations of this procedure which are now at the point of providing  a semi-quantitative  solution for the normal-state \cite{Gull10_clustercompare} and superconducting \cite{Gull12} properties the two-dimensional square-lattice Hubbard model, which is widely believed to capture the essential physics of the high transition-temperature superconductors. Crucially, the new methods enable access to  clusters large enough to provide  confidence that the true behavior of the Hubbard model is revealed, and to low enough temperatures that the superconducting state can be constructed. 

Dynamical mean field methods have determined that the Hubbard model exhibits a ``pseudogap'' \cite{Huscroft01,Parcollet04,Civelli05,Civelli08,Macridin06,Werner098site,Gull09,Ferrero09,Ferrero09B,Ferrero10,Liebsch08,Liebsch09,Sakai09,Sakai10,Lin10,Gull10_clustercompare,Sordi10,Sordi11,Sordi12,Sordi13} where some regions of the Brillouin zone are gapped and others are not \cite{Werner098site,Gull09,Ferrero09,Ferrero09B} as well as a $d_{x^2-y^2}$-symmetry superconducting state \cite{Lichtenstein00,Maier00,Maier05,Yang11,Gull12}. The superconducting state can now be constructed and aspects of its interplay with the pseudogap can be studied \cite{Gull12}. Quasiparticle properties \cite{Gull12} and   energetics \cite{Gull12b} have been determined. In this paper we use the methods to study the superconductivity-induced changes in the interplane conductivity and $B_{1g}$ Raman scattering cross section of the two-dimensional Hubbard model. 

The rest of the paper is organized as follows. Section \ref{Introduction} presents the response functions to be computed, gives more specifics of the physical phenomena of interest and outlines the theoretical methods used. Section \ref{Results} presents the main new results of the paper: a computation of the interplane conductivity and Raman scattering cross sections. Section \ref{Discussion} analyses the results and their relation to experiment. Section \ref{Conclusions} summarizes the findings and outlines future directions for research. Appendices provide calculational details. 

\section{Introduction \label{Introduction}}
\subsection{Model} The essential structural motif of the high transition temperature copper oxide superconductors is the CuO$_2$ plane, a square planar array of Cu ions, with an oxygen ion at the midpoint of each Cu-Cu bond. It is by now accepted that the interplane coupling is weak enough that it may for most purposes be neglected, so that the basic physics problem which must be understood concerns the motion of electrons in a two dimensional lattice with a square symmetry.   The interplane conductivity may then be studied by second order perturbation theory in the interplane coupling. 

In the ``parent compounds'' of high-$T_c$ superconductors, the density of electrons is one per CuO$_2$ unit, but the materials are insulating with a large ($\sim 1.5-2$ $eV$) band gap \cite{Uchida91}. The insulating behavior is widely supposed to be a consequence of strong electronic correlations, related to the ``Mott insulating'' phenomenon \cite{Anderson87}. Removing electrons (adding holes) produces metallic and superconducting behavior. The superconducting state is of $d_{x^2-y^2}$ symmetry \cite{Wollman93}, with the superconducting gap being maximal at the center of  the zone face  ( $(\pi,0)$ ) and vanishing along the zone diagonal ($(0,0)\rightarrow(\pi,\pi)$) direction. Adding holes also produces a region of ``pseudogapped'' behavior \cite{Alloul89,Ding96,Huefner08}, in which even at temperatures above the superconducting transition temperature the electronic density of states is suppressed in the zone face region but not in the zone diagonal region. 

A widely, but not universally, accepted hypothesis \cite{Anderson87} is that the basic theoretical model which describes the physics of interest is the two dimensional one-orbital square lattice Hubbard model, an idealized description which arises as a low ($\omega \lesssim 1.5eV$) energy effective model of the underlying material Hamiltonian. This model represents the physical situation in terms of electrons moving among sites of a two dimensional square lattice, and subject to an interaction $U$ which disfavors double occupancy. In a mixed position/momentum representation we have
\begin{equation}
H=\sum_{k\sigma}\varepsilon_kc^\dagger_{k\sigma}c_{k\sigma}+U\sum_in_{i\uparrow}n_{i\downarrow}.
\label{Hhub}
\end{equation}
Here $c^\dagger_{i\sigma}$ creates an electron of spin $\sigma=\uparrow,\downarrow$ on site $i$ of a two dimensional square lattice of unit lattice constant, $c^\dagger_{k\sigma}$ is its Fourier transform to momentum space, $\varepsilon_k$ is the energy dispersion and  $n_{i\uparrow}=c^\dagger_{i\uparrow}c_{i\uparrow}$ is the operator density of up-spin electrons on site $i$.   In the computations presented below we take  $\varepsilon_k=-2t\left(\cos~k_x+\cos~k_y\right)$ with $t=0.35eV$. While the dispersion is particle-hole symmetric we consider non-zero dopings for which the particle-hole symmetry is broken. 

Because we wish to treat superconducting phenomena it will be convenient to write subsequent equations in terms of the Nambu spinors
\begin{equation}
\Psi^\dagger_k=\left(\begin{array}{c}c^\dagger_{k\uparrow} \\c_{-k\downarrow}\end{array}\right)
\label{psidef}
\end{equation}
and the corresponding matrix Nambu Green function defined for imaginary time $\tau>0$ as 
\begin{equation}
\mathbf{G}(k,\tau)=-\left\langle\Psi^\dagger_k(\tau)\Psi_k(0)\right\rangle.
\label{GNambu}
\end{equation}

%In this paper we aim to determine how the interplane conductivity and $B_{1g}$ Raman response of the two dimensional Hubbard model relate to measurements of these quantities on high-$T_c$ copper oxide superconducting materials.  The Raman cross section is computed directly from the  two dimensional Hubbard model; the interplane conductivity is computed using leading order perturbation theory in the coupling in the third dimension. 

\subsection{Formalism: dynamical mean field method}
Evaluation of the interplane conductivity and Raman scattering amplitude require knowledge of the normal and anomalous components of the electron Green function,  Eq.~\ref{GNambu}. We obtain these  using the dynamical cluster approximation (DCA) \cite{Hettler98,Maier05} version  of cluster dynamical mean field theory.  In this method the Brillouin zone is divided into some number $N$ of equal volume patches labeled by a central momentum  $K$ and the electron self energy $\Sigma(k,\omega)$ is  represented as a piecewise constant function taking different values in each patch, so defining $\phi_K(k)=1$ if $k$ is in the patch labeled by $K$ and $\phi_K(k)=0$ otherwise, 
\begin{equation}
\mathbf{\Sigma}(k,\omega)=\sum_K\phi_K(k)\mathbf{\Sigma}_K(\omega).
\label{sigmadef}
\end{equation}
The $\Sigma_K$ are obtained from the solution of an N-site quantum impurity model specified by the original Hubbard interaction and a self-consistency equation which relates lattice $(G(k,\omega))$ and  impurity $(G_K(\omega))$ quantities:
\begin{equation}
\mathbf{G}_{K}(\omega)=N\int \frac{d^2k}{(2\pi)^2}\phi_K(k)\mathbf{G}(k,\omega).
\label{SCE}
\end{equation}

The solution of the impurity model in the normal and superconducting states is obtained using continuous-time quantum Monte-Carlo methods \cite{Gull11} with submatrix updates \cite{Gull08,Gull10_submatrix} as described in detail in Appendix \ref{Numerics}.  These methods provide solutions on the imaginary time or Matsubara frequency axis; real frequency information is obtained from maximum entropy analytical continuation as described in Appendix \ref{Continuations}. 

\begin{figure}[htb]
\includegraphics[width=0.95\columnwidth]{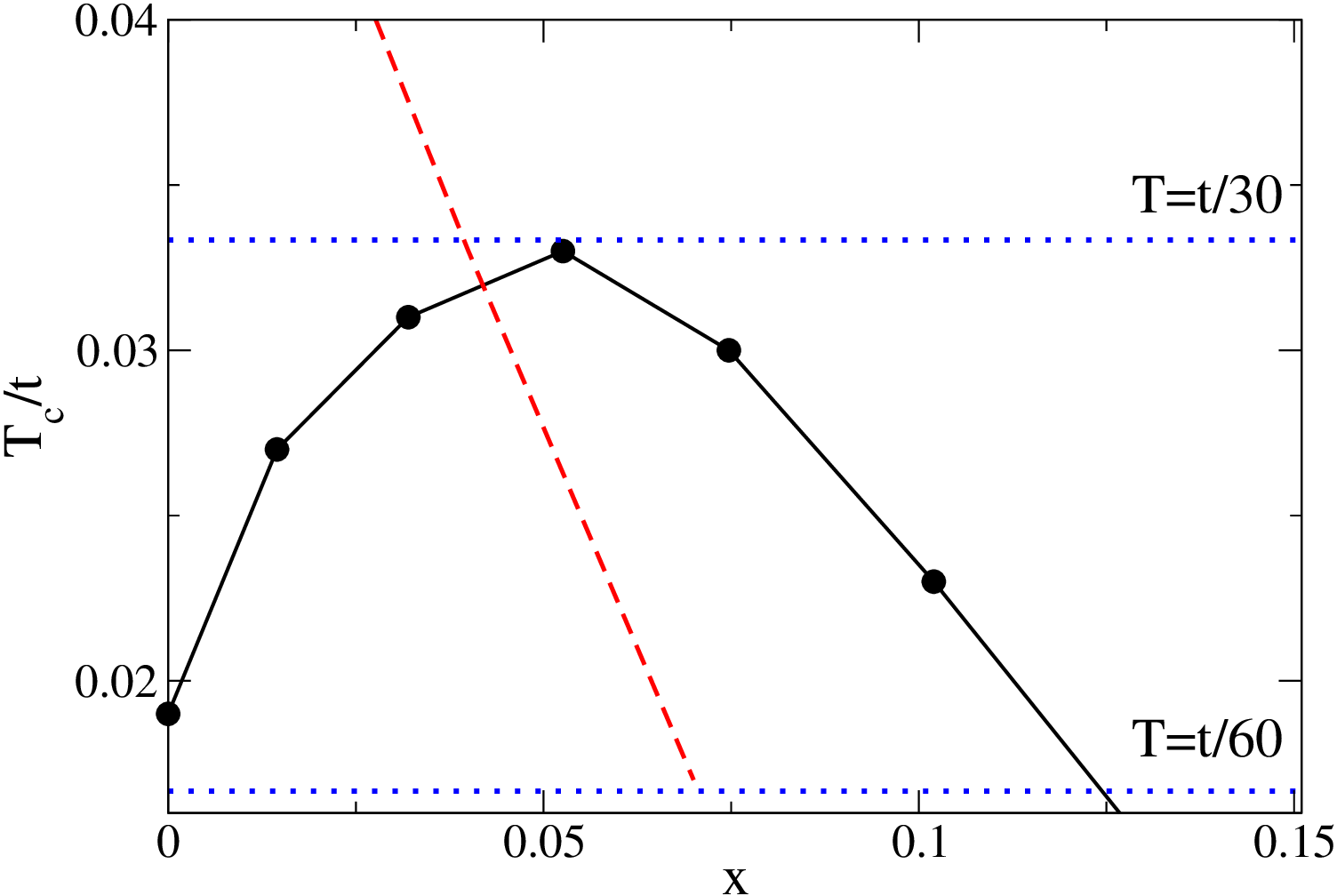}\caption{ Solid lines and filled circles (black online): superconducting critical temperature of the Hubbard model with nearest neighbor hopping calculated for $U=6t$ using the 8-site dynamical cluster approximation (see also Ref.~\onlinecite{Gull12}). Dashed line (red online) denotes crossover to normal state pseudogap. Dotted lines (blue online)  indicate the temperatures studied in this paper. Note that the temperature axis is cut off slightly below the lowest temperature used in this study. 
}\label{fig:Tc}
\end{figure}

The computational burden of DCA calculations rises rapidly with cluster size $N$ and interaction strength $U$. With the computational resources available to us, study of temperatures below the superconducting transition temperature at generic dopings  is feasible for $U\lesssim 7$ at $N=8$ but to obtain data of the precision needed for analytical continuation of response functions we employ $U=6$. Fig.~\ref{fig:Tc} plots the transition temperature against carrier concentration for this $U$. The onset of the normal state pseudogap is also indicated as a dashed line. As will be discussed in the conclusions this $U$ probably is slightly lower than the $U$ which is relevant to the real materials. In consequence the superconducting region is pushed to lower carrier concentrations than observed in the real high$T_c$ materials. 

\subsection{Raman  $B_{1g}$ and Raman  $B_{2g}$ cross section}
Raman scattering is a process in which an incident photon at some frequency $\omega_{in}$ polarized along some direction $a$ is scattered to an outgoing photon at some other frequency $\omega_{out}=\omega_{in}-\Omega$  and some other polarization direction $b$. The details of the Raman scattering process are complicated. Both phonons and electrons may be important. The different choices of incident and outgoing polarization lead to many different symmetry channels. Often, to enhance the signal, the incident or outgoing photon frequencies are tuned to be near resonance with some other excitation of the solid. Providing a detailed treatment of all of these effects is beyond the scope of this paper. Here we will focus on electronic scattering in the ``$B_{1g}$'' channel  at moderate energy transfers ($\Omega$ small compared to bandwidths or interband transition energies). In this case the basic Raman process is the creation of a particle-hole pair of negligible net momentum. For  $B_{1g}$ symmetry the  amplitude for this process vanishes for electrons along the zone diagonal and is maximal for electrons at the zone face. We will also present a few results for   $B_{2g}$ symmetry, where the amplitude is maximal for zone-diagonal electrons and vanishes at the antinodes.  We follow the convention in the literature and assume the simplest forms compatible with the desired symmetries, which are
\begin{equation}
H_\text{Raman}=\sum_{k\sigma}\mathcal{R}_kc^\dagger_{k,\sigma}c_{k\sigma}
\label{HRaman}
\end{equation}
with (in $B_{1g}$ geometry)
\begin{equation}
\mathcal{R}_k^{B_{1g}}=\mathcal{R}\left(\cos k_x-\cos k_y\right)
\label{Rkdef}
\end{equation}
and (in $B_{2g}$ geometry)
\begin{equation}
\mathcal{R}_k^{B_{2g}}=\mathcal{R}\sin k_x\sin k_y.
\label{Rkdef2}
\end{equation}

The amplitude $\mathcal{R}$  is not important for our purposes. Eqs.~\ref{Rkdef},\ref{Rkdef2} may be derived in the non-resonant limit by use of a minimal coupling ansatz and are widely used in the literature. 

Standard linear response methods may then be used to obtain the susceptibility $\chi(\Omega)$ which characterizes the Raman response. The result involves both the electron Green function and a vertex function describing the interaction of the particle-hole pairs created by the Raman process. Computation of the vertex function is very challenging. With the methods used here it has been recently attempted for a wide frequency range analysis of the normal state Raman cross section \cite{Lin12}. The method is too expensive to be run in the superconducting state on today's computers but the results of Ref.~\onlinecite{Lin12} indicate that while vertex corrections have a crucial effect at low dopings and high frequencies ($\sim 0.5eV$ where they express e.g. the two-magnon contribution) they are much less important at the low frequencies of interest here.  We therefore neglect the vertex corrections, so that on the Matsubara axis and in Nambu notation
\begin{equation}
\chi(\Omega_n)=T\sum_{\omega_n,k}\text{Tr}\left[\mathcal{R}^2_k\tau_3\mathbf{G}(k,\omega_n+\Omega_n)\tau_3\mathbf{G}(k,\omega_n)\right].
\label{chiraman}
\end{equation}
The presence of the $\tau_3$ factors in the expression for the  Raman reflects the fact that the Raman probe does couple to a quasiparticle at the gap edge.  We evaluate the response function on the imaginary axis and then analytically continue the result (see appendix \ref{Continuations}).

\subsection{Interplane conductivity}
The computation of the interplane coupling begins from an ansatz for the coupling between planes. Electronic structure calculations indicate that the coupling is small,\cite{Andersen94,Lichtenstein96,Novikov93,Ioffe98} so only electron hopping between adjacent planes need be considered.  This is parametrized by a hopping amplitude $t_\perp(k_x,k_y)$ which connects an electronic state at in-plane momentum $(k_x,k_y)$ in one plane with an electronic state of the same in-plane momentum in an adjacent plane. (Some authors have considered impurity-mediated interplane coupling \cite{Kim98} which does not conserve momentum, but we will restrict attention to the ideal undisordered situation here). Gauge invariance considerations then imply that the coupling of electrons to an electric field in the interplane direction may be determined by multiplying the interplane hopping by the usual Peierls phase factor involving the vector potential $\vec{A}=A\ez$ with $\ez$ denoting the direction perpendicular to the planes.  Band theory and tight binding \cite{Chakravarty98,Ioffe00} considerations indicate that $t_\perp$ has a strong momentum dependence for simple ``one-layer'' systems such as La$_{2-x}$CuO$_4$ or TlBa$_2$CuO$_6$  so the final result is (setting $e=\hbar=c=$interplane distance$=1$)
\begin{equation}
t_\perp(k_x,k_y;A)=-t_\perp\left(\cos k_x-\cos k_y\right)^2e^{iA}.
\label{tperpform}
\end{equation}
Thus the interplane hopping amplitude vanishes for momenta along the diagonals of the two dimensional Brillouin zone and is maximal at the zone faces implying, as many authors have noted\cite{Chakravarty98,Kim98,Ioffe00}, that the c-axis conductivity is in effect a spectroscopy of the behavior of the zone-face electrons. In bilayer systems such as YBa$_2$Cu$_3$O$_{6+x}$ the inter-bilayer coupling has the form given by Eq.~\ref{tperpform} but the intra-bilayer coupling may also have contributions from the zone diagonal electrons \cite{Dubroka10}. These complications will not be addressed in this paper. 

In Nambu notation the Hamiltonian giving the interplane coupling is then 
\begin{equation}
H_\perp=\sum_{j;k_x,k_y\sigma}\Psi^\dagger_{j+1,kx,ky}t_\perp(k_x,k_y)e^{iA\tau_3}\tau_3\Psi_{j,kx,ky}
+H.c..\label{Hperpdef}
\end{equation}
Here $\tau_3$ is the Pauli matrix acting in Nambu space and for simplicity we have assumed $t_\perp$ to be the same between all planes. 

We now use standard linear response theory to write an expression for the conductivity valid to leading nontrivial order in both the applied vector potential and the interplane coupling. The conductivity is  as usual the sum of paramagnetic and diamagnetic terms. In Matsubara space we have 
\begin{equation}
\sigma_c(\Omega_n)=\sigma_c^D(\Omega_n)+\sigma_c^P(\Omega_n)
\label{sigmaDbasic_A}
\end{equation}
with 
\begin{equation}
\sigma_c^D(\Omega_n)=\frac{T}{\Omega_n}\sum_{mk}\text{Tr}\left[t_\perp(k)^2\tau_3\mathbf{G}(k,\omega_m)\tau_3\mathbf{G}(k,\omega_m)\right]
\label{sigmaD}
\end{equation}
\begin{equation}
\sigma_c^P(\Omega_n)=\frac{T}{\Omega_n}\sum_{mk}\text{Tr}\left[t_\perp(k)^2\mathbf{G}(k,\omega_m+\Omega_n)\mathbf{G}(k,\omega_m)\right].
\label{sigmaP}
\end{equation}
The absence of $\tau_3$ factors in the expression for $\sigma_c^P$  encodes the fact that a quasiparticle with energy equal to the superconducting gap is an equal admixture of electron and hole states, which because it is chargeless does not couple to an applied electric field. In materials with a bilayer structure  extra contributions appear in the conductivity related to intrabilayer plasmon excitations \cite{Dubroka10}; these will not be considered here. 

In one important respect the piecewise constant self energy,  an essential feature of the  DCA dynamical mean field approximation, is problematic. As can be seen from Eq.~\ref{tperpform} there is an interplay between the magnitude of the superconducting gap $\sim |k_x-k_y|$ and the interplane coupling. Near the nodes the superconducting gap is larger than the interplane coupling, meaning that the contribution of nodal quasiparticles to the low frequency response is strongly suppressed. On the other hand, in the DCA approximation the entire momentum sector containing the nodal point is gapless, so that the Drude response of  nodal quasiparticles contribute strongly to the low frequency  response. In our calculations we therefore suppress the nodal region completely, by integrating only over momenta corresponding to the antinodal sector.  (This issue also arises in the Raman case, but is not important for the analysis because the response is suppressed at low frequencies). One may view our approximation as using an interlayer coupling which has the same momentum discretization as the DCA approximation. 

The  c-axis superfluid stiffness $\rho_{c,S}$ is given by \cite{Schrieffer99}
\begin{eqnarray}
\rho_{c,S}&=&T\sum_{n,k}t_\perp(k)^2\text{Tr}\bigg[\tau_3\mathbf{G}(k,\omega_n)\tau_3\mathbf{G}(k,\omega_n)
\nonumber \\
&-&\mathbf{G}(k,\omega_n)\mathbf{G}(k,\omega_n)\bigg]
\label{rhos}
\end{eqnarray}
and the   interplane conductivity may be rewritten as
\begin{equation}
\sigma_c(\Omega_n)=\frac{\rho_{c,S}}{i\Omega_n}+\sigma_c'(\Omega_n)
\label{sigmacfinal_matsu}
\end{equation}
with 
\begin{equation}
\sigma'_c(\Omega_n)=\sigma^P_c(\Omega_n)-\frac{T}{\Omega}\sum_{n,k}t_\perp(k)^2
 \text{Tr}\left[\mathbf{G}(k,\omega_n).\mathbf{G}(k,\omega_n)\right].
\label{sigmacprime}
\end{equation}

To obtain real-frequency conductivities we construct $\chi_c(\Omega_n)=\Omega_n\sigma_c^P(\Omega_n)$, which is then analytically continued using the methods employed for the Raman case as discussed  in Appendix \ref{Continuations} and from this construct the continued $\sigma_c^{'}$ and thus $\sigma_c$.

We remark that the interplane spectral weight 
\begin{equation}
S_c=\int \frac{d\omega}{\pi} \sigma_c(\omega)
\label{Scdef}
\end{equation}
is given by $\Omega_n\sigma_c^D(\Omega_n)$ \cite{Ioffe00} (note that the integral includes the superfluid delta function if present).

\section{Results \label{Results}}
\subsection{Raman  $B_{1g}$ and Raman  $B_{2g}$ cross section}
%\subsection{Raman Cross section}
\begin{figure*}[htb]
\includegraphics[width=0.95\textwidth]{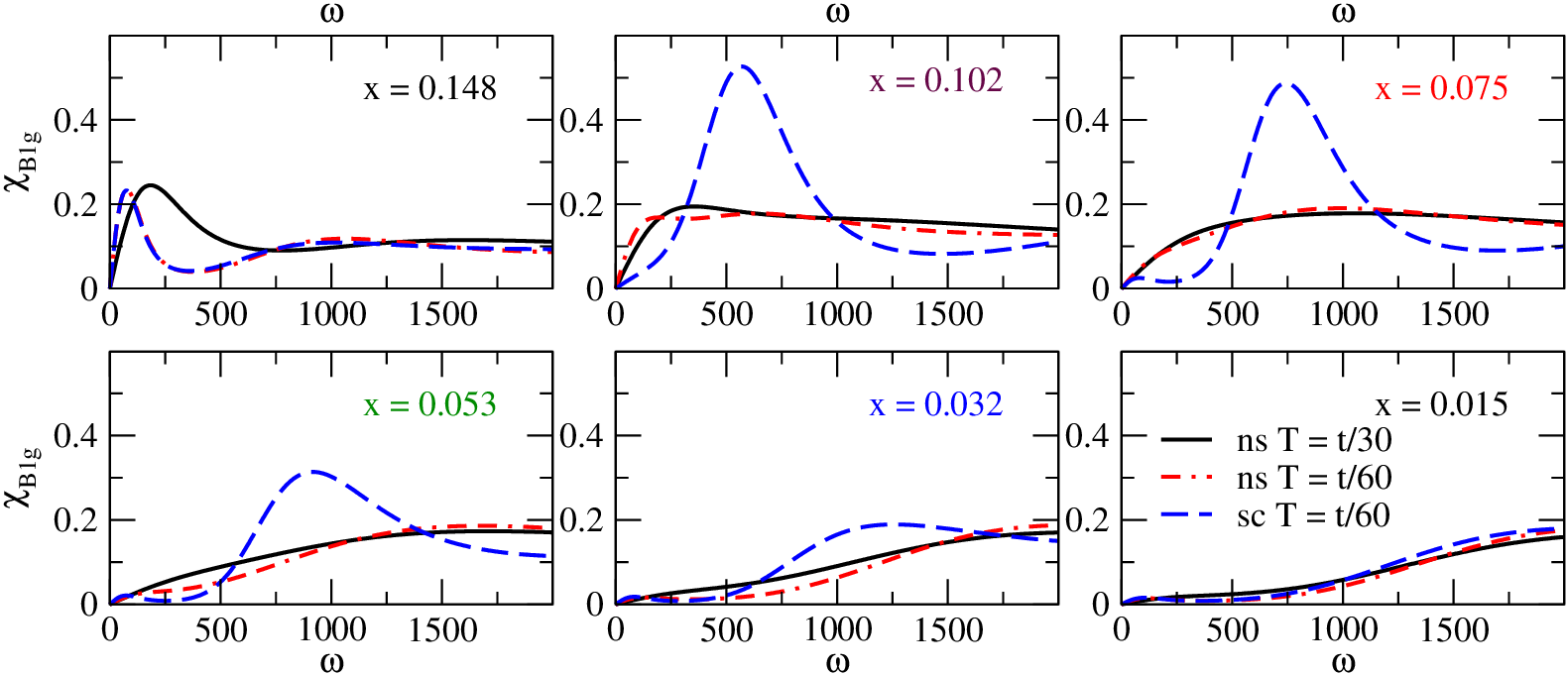}\caption{ $B_{1g}$ Raman scattering cross section plotted against frequency (expressed in wave numbers using $t=0.35eV$)  for the two dimensional Hubbard model with $U=6t$ at carrier concentrations 0.148 (top left), 0.102 (top middle),  0.075 (top right), 0.053 (bottom left), 0.032 (bottom middle) and  0.015 (bottom right). Solid curve (black online) $T=t/30$; dot-dashed curve (red online) $T=t/60$, normal state; dashed curve (blue online) $T=t/60$ superconducting state.  Label shading (color online) corresponds to color of curves for this doping in other figures.
}
\label{fig:raman}
\end{figure*}
Figure   ~\ref{fig:raman}  presents one of the  principal results of this paper: the doping dependence of the  $B_{1g}$-symmetry Raman response  of the two dimensional Hubbard model for several carrier concentrations at interaction strength $U=6t$.  This sequence of carrier concentrations corresponds to the cuts  across the phase diagram shown as dotted lines in Fig.~\ref{fig:Tc}. The cuts  include both Fermi liquid and pseudogapped regimes, and temperatures both above $T=t/30\approx 140K$ (using $t=0.35eV$) and well below the highest superconducting transition temperatures ($T=t/60\approx 70K \approx T_c^{max}/2$). 

The upper left panel of  Figure~\ref{fig:raman} presents results obtained for a high doping just outside the regime where superconductivity is found at the temperatures we have studied. The high temperature Raman scattering amplitude (solid curve, black online) has the features expected of a strongly correlated Fermi liquid: a low frequency peak (visible at $\omega\approx 250 ~cm^{-1}$) is characteristic of coherently propagating quasiparticles, while the relatively featureless higher frequency scattering intensity is attributable to the incoherent part of the electron spectral function. The peak frequency is related to the quasiparticle scattering rate.  Indeed for quasiparticles with a momentum non-conserving scattering rate $\Gamma_{qp}/2$ the Raman cross section is 
\begin{equation}
\chi^{qp}_{B_{1g}}(\Omega)=\chi_0\frac{\frac{\Omega}{\Gamma_{qp}}}{1+\left(\frac{\Omega}{\Gamma_{qp}}\right)^2}.
\label{chiqp}
\end{equation}
As the temperature is decreased the low frequency peak moves to lower energy, indicating a decrease in scattering rate.  For further discussion of the normal state issues see Refs.~\onlinecite{Lin10,Lin12} and note that especially at lower dopings the frequency dependence at frequencies $\gtrsim 1500 cm^{-1}$ is affected by vertex corrections not considered here \cite{Lin12}.  

The upper middle and upper right  panels of Figure~\ref{fig:raman}  show dopings  within the superconducting regime. For $x=0.1$  the normal-state Fermi liquid coherence effects are almost invisible. At the higher temperature $T=t/30$ the scattering rate is large enough that the quasiparticle peak has merged with the continuum.  As the temperature is lowered in the normal state a modest increase in the low frequency slope (decrease in scattering rate) is evident and a hint of a quasiparticle peak is seen as a very weak minimum at $\Omega\sim250cm^{-1}$, but there is no clear signature of coherent Fermi liquid  behavior  down to our lowest accessible temperature.  $x=0.075$ is a carrier concentration at the boundary of the pseudogap regime and near the maximum in the $T_c$ vs doping plot. The normal state traces show a larger scattering rate (lower slope at  $\Omega\rightarrow 0$) and no trace of quasiparticle coherence is visible. 

At these carrier concentrations the  onset of superconductivity has dramatic effects. The low frequency intensity is suppressed, and a large peak becomes evident. This peak arises from quasiparticle excitations across the superconducting gap. It is visible in this response function because the coherence factors are such that a quasiparticle at the gap edge couples to the Raman probe. In an  s-wave superconductor with a uniform single-particle gap $\Delta$ the Raman intensity would diverge $\sim 1/\sqrt{\Omega-2\Delta}$. In a d-wave superconductor the momentum dependence of the gap eliminates the divergence, leaving just a peak, but in  our calculation the piecewise constant nature of the self energy means that a  divergence similar to the s-wave case will occur.  However, the relatively strong inelastic scattering leads to substantial thermal  broadening of the divergence at the temperatures accessible to us. The peak structure still provides a good estimate of the gap $2\Delta$.   At higher frequencies the superconducting state Raman amplitude appears to be smaller than the normal state amplitude, but these differences are at the edge of what can  reasonably be resolved with presently available analytical continuation techniques so it is not clear how much significance can be attributed to this difference. 

\begin{figure}[tb]
\includegraphics[width=0.95\columnwidth]{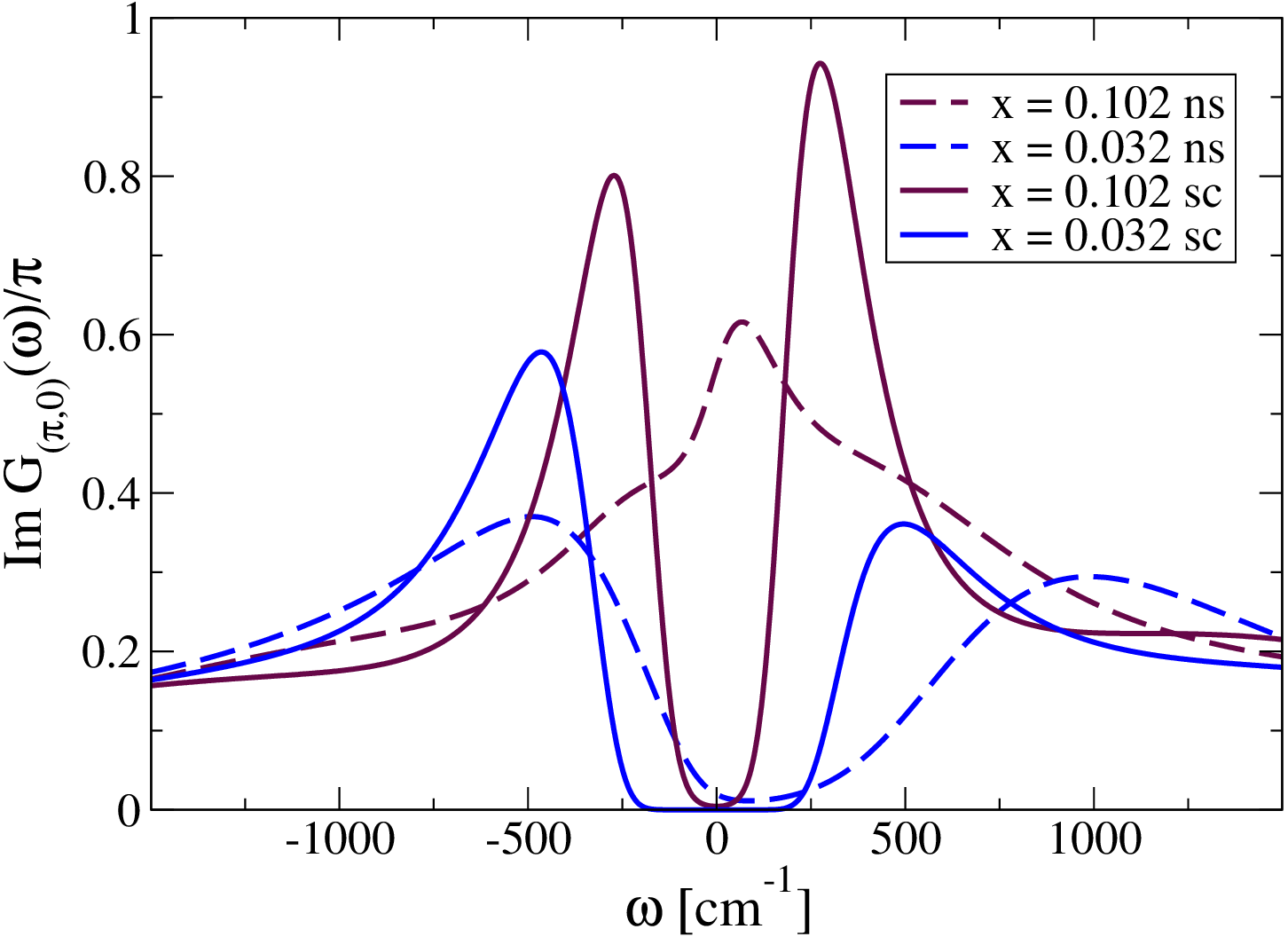}
\caption{Imaginary part of $(0,\pi)$ sector electron Green function (divided by $\pi$) in normal (dashed lines) and superconducting (solid lines) states for doping $x=0.102$ (heavy lines, maroon online) and $x=0.032$ (lighter lines, blue online).}
\label{fig:spectra}
\end{figure}
The association between the peak in the Raman scattering intensity and the superconducting gap may be verified from an examination of the electron spectral function (imaginary part of the electron Green function divided by $\pi$) shown for two carrier concentrations in Fig.~\ref{fig:spectra}. There is not a one-to-one correspondence, since the Green function is averaged over the $(0,\pi)$ momentum sector while the Raman intensity is in essence the average of a product of two $G$ over the sector, and  also involves coherence factors, but one may expect energy scales to be similar. The lighter weight lines (maroon  online) show the normal (dashed line) and superconducting (solid line) spectral functions for the doping $x=0.102$ whose Raman intensity is displayed in the upper middle panel of Fig.~\ref{fig:raman}. No normal state pseudogap is visible. In the superconducting state a clear gap is evident. The peak-to-peak separation $\sim 600~cm^{-1}$ is seen to coincide with the peak position in the Raman intensity.

The left and middle panels of the lower row of Fig.~\ref{fig:raman} display results for lower carrier concentrations, now well within the pseudogap regime.  The pseudogap can be identified from the decrease in low frequency normal-state Raman intensity as temperature is lowered and a normal state pseudogap scale can be approximately identified from the frequency at which the lower temperature normal state trace rejoins the high temperature trace ($\sim 1100~cm^{-1}$ for $x=0.053$ and $\sim 1400~cm^{-1}$ for $x=0.032$). These estimates are in reasonable accord with the pseudogap scales inferred from the normal state Green function. For example, the heavier dashed lines (blue online) in Fig.~\ref{fig:spectra} show the normal state Green function for doping $x=0.032$. The pseudogap is clearly visible in the normal state Green function and defining the pseudogap energy scale $2\Delta_{pg}$ as the peak to peak separation in the spectral function gives $2\Delta_{pg}\approx 1400~cm^{-1}$ in agreement with the estimate from the Raman spectrum. 

\begin{figure}[tb]
\includegraphics[width=0.95\columnwidth]{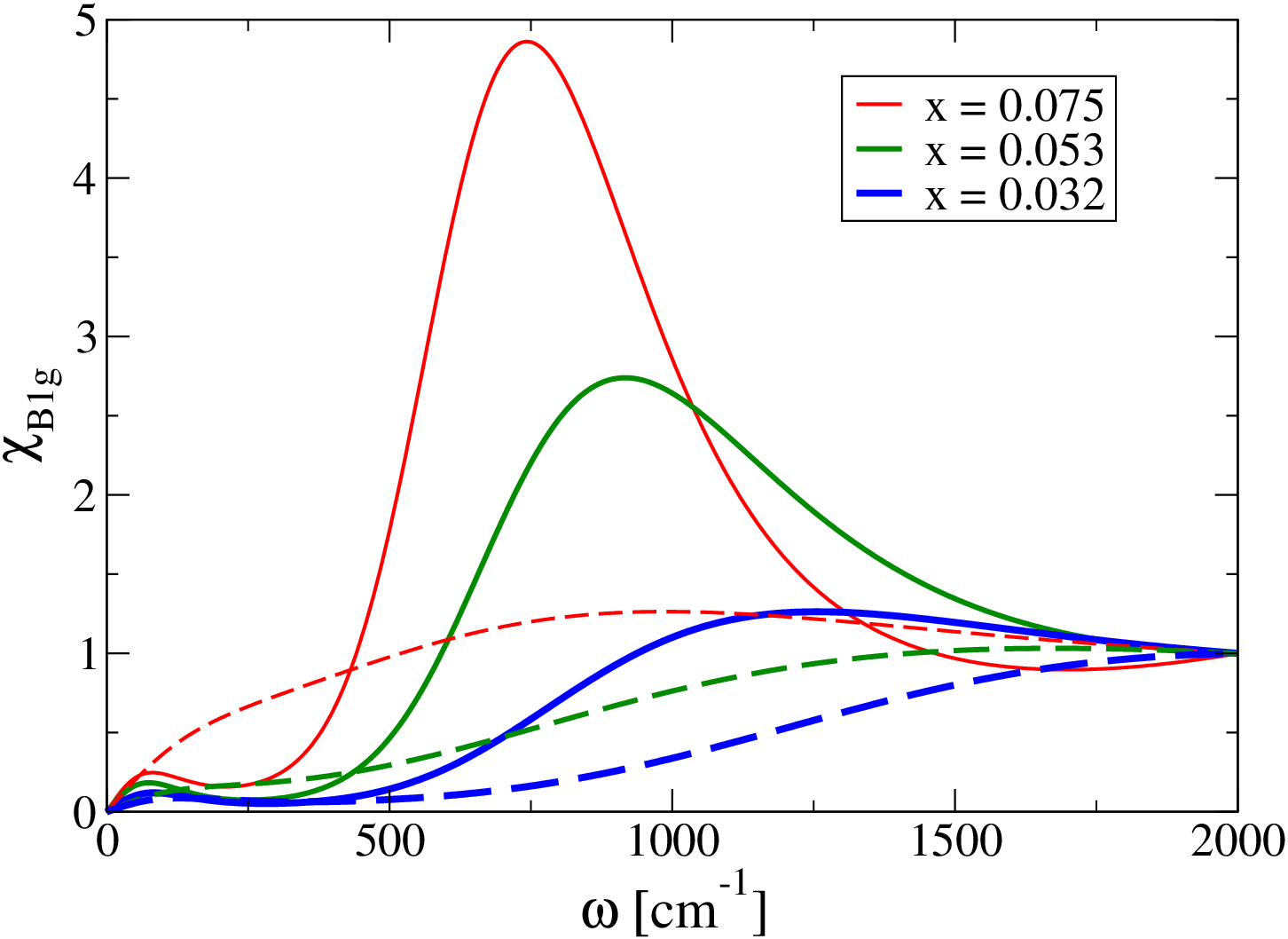}\caption{ $B_{1g}$ Raman scattering cross section in the normal (dashed) and superconducting (solid) state, for dopings $x = 0.075$ (light lines, red online), $x = 0.053$ (intermediate lines, green online) and $x = 0.032$ (heavy lines, blue online). All curves are normalized to the value at $2000~cm^{-1}$.  
}
\label{fig:ramanns}
\end{figure}

For these dopings superconductivity again produces a peak in the Raman cross section, but the peak is broader, and the relative increase in amplitude less, than at higher dopings. To demonstrate this more clearly, in Fig.~\ref{fig:ramanns} we present the normal and superconducting state $B_{1g}$ Raman scattering intensity, normalized  to the value at $2000~cm^{-1}$ for $x=0.102$ (intermediate shading, maroon online) $0.053$ (light shading, red online) and $0.032$ (dark shading, blue online).   This rescaled plot makes it clear that as parameters are tuned through the pseudogap phase to the low doping superconducting boundary the change in Raman intensity due to   superconductivity  weakens both in relative and in absolute terms. 

\begin{figure}[tb]
\includegraphics[width=0.95\columnwidth]{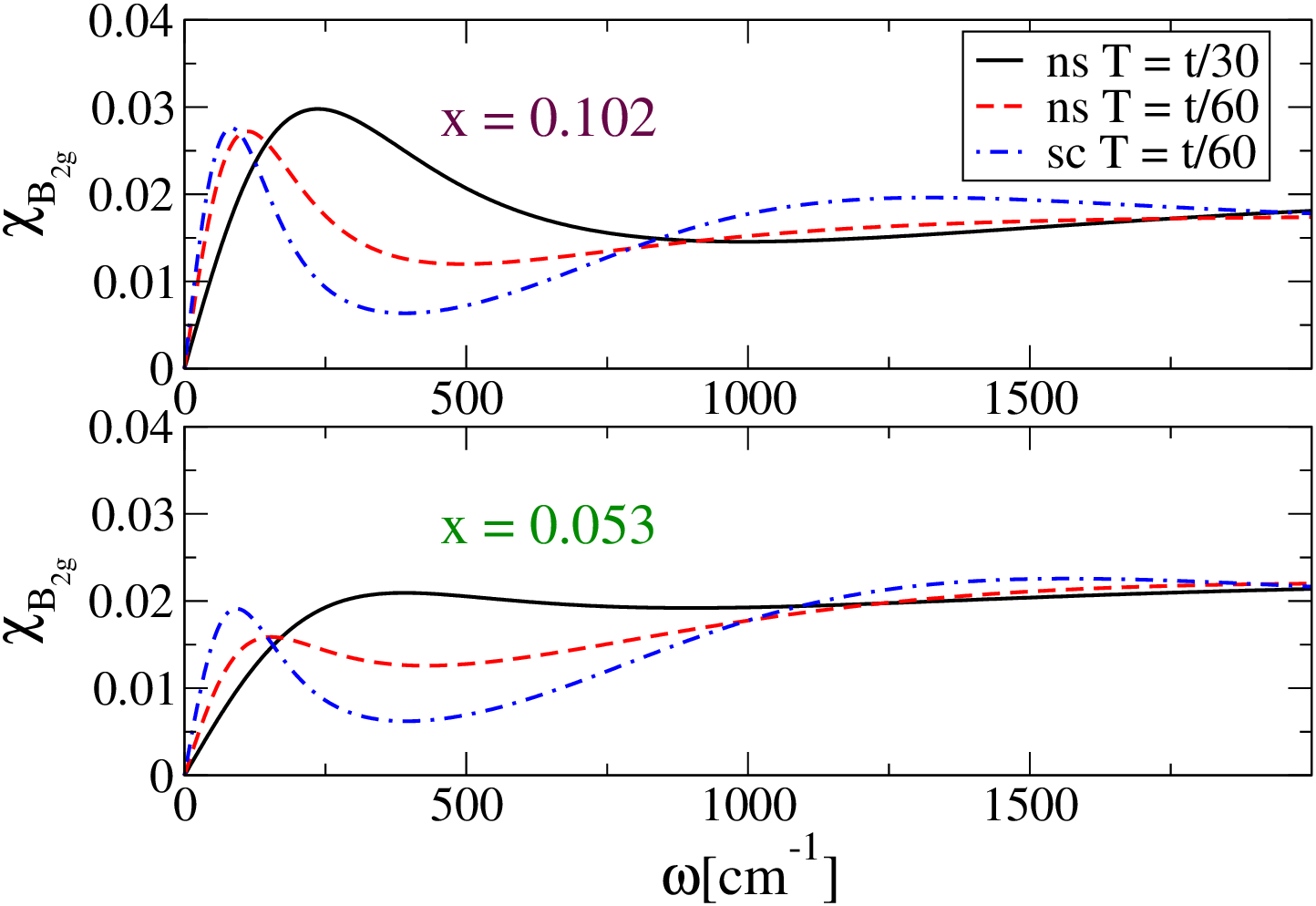}\caption{ $B_{2g}$ Raman scattering cross section in the normal state at high $T$ (solid line, black online) and at low $T$ (dashed line, red online) and in the superconducting state at low $T$ (dot-dashed, blue online), for carrier concentration $x=0.102$ (top panel) and $0.053$ (bottom panel).  
}
\label{fig:ramanB2G}
\end{figure}

We have also computed the Raman scattering cross section in the $B_{2g}$ scattering channel. The $B_{2g}$ matrix element (Eq.~\ref{Rkdef2}) is maximal in the zone diagonal sector, and contributions from the antinodal sector which carries the information about pseudogap and superconductivity are small. Representative results are displayed in  Fig.~\ref{fig:ramanB2G} The evolution of the spectra as temperature is lowered in the normal state is consistent with that reported previously\cite{Lin10}. At all dopings a clear quasiparticle peak is observed which becomes better defined as temperature is lowered.  The difference with the normal state $B_{1g}$ spectra reflects the momentum-space differentiation characteristic \cite{Ferrero09,Gull10_clustercompare} of the approach to the Mott transition in this model. The zone diagonal states which dominate the $B_{2g}$ response remain more or less Fermi liquid-like while more exotic physics affects the states near the zone face which determine the $B_{1g}$ response. The effects of superconductivity are  difficult to discern in the calculated spectrum because in the DCA approximation used here the piecewise constant nature of the self energy means that the anomalous self energy vanishes in the entire zone-diagonal momentum sector where the $B_{2g}$ matrix element is peaked.  A DCA calculation of the effects of superconductivity on the $B_{2g}$ spectra would require a much finer momentum space resolution, corresponding to $N=32$ or larger,  to capture the behavior of the superconducting gap near the  nodes. These cluster sizes are not cannot yet be studied in the relevant range of T and U.

\subsection{Interplane conductivity}
\begin{figure*}[tb]
\includegraphics[width=0.95\textwidth]{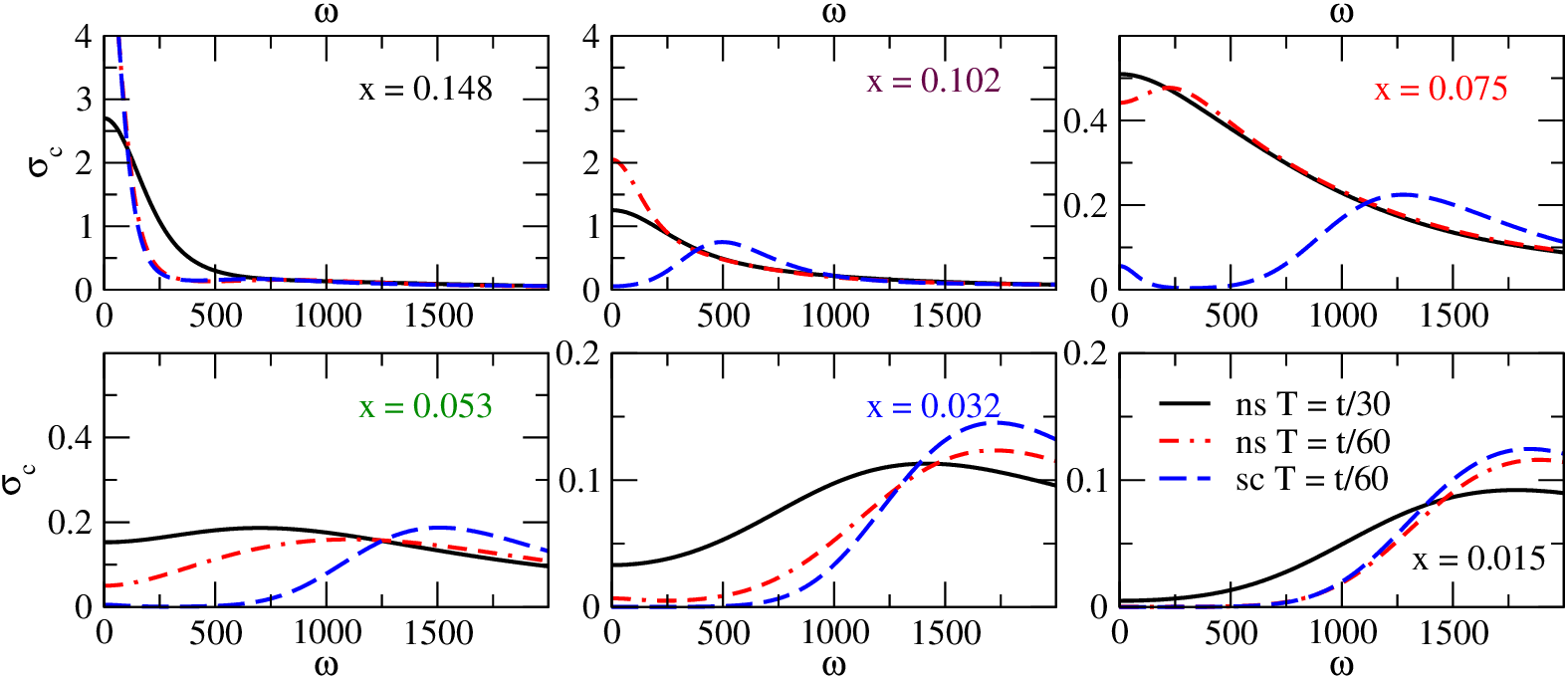}\caption{Real part of interplane conductivity plotted against frequency (expressed in wave numbers using $t=0.35eV$)  for the two dimensional Hubbard model with $U=6t$ at carrier concentrations 0.148 (top left), 0.102 (top right),  0.075 (middle left), 0.053 (middle right), 0.032 (bottom left) and  0.015 (bottom right). Solid curve (black online) $T=t/30$; dot-dashed curve (red online) $T=t/60$, normal state; dashed curve (blue online) $T=t/60$ superconducting state $(n>0.02)$. Label colors (online)  distinguish doping levels: in other figures different dopings are indicated by the corresponding colors.  }
\label{fig:caxis}
\end{figure*}
Fig.~\ref{fig:caxis} shows the interplane conductivity calculated for the same parameters as the Raman scattering shown in Fig.~\ref{fig:raman}.  The normal state physics has previously been discussed in the context of 8-site DMFT calculations \cite{Lin10}  and similar results for $N=2$ and $N=4$ have also been presented.\cite{Ferrero09,Sordi13} Our normal state results are consistent with these previous works. The overdoped case (upper left panel)  exhibits a well defined low frequency ``Drude'' peak which rapidly sharpens as the temperature is decreased. As the doping is decreased first the Drude peak is broadened and decreased in amplitude, then the temperature dependence ceases and a hint of pseudogap becomes visible (upper right panel). With further decrease of doping the pseudogap becomes obvious; the pseudogap magnitude increases as the doping decreases. Note the large changes in $y$-axis scale between the two first panels, the panels for intermediate doping, and the panels for low doping, reflecting the drastic reduction of low-frequency  interplane conductivity as doping is decreased although for frequencies greater than $\gtrsim 3000cm^{-1}$ the calculated interplane conductivity has only a weak doping dependence.

\begin{figure}[tb]
\includegraphics[width=0.95\columnwidth]{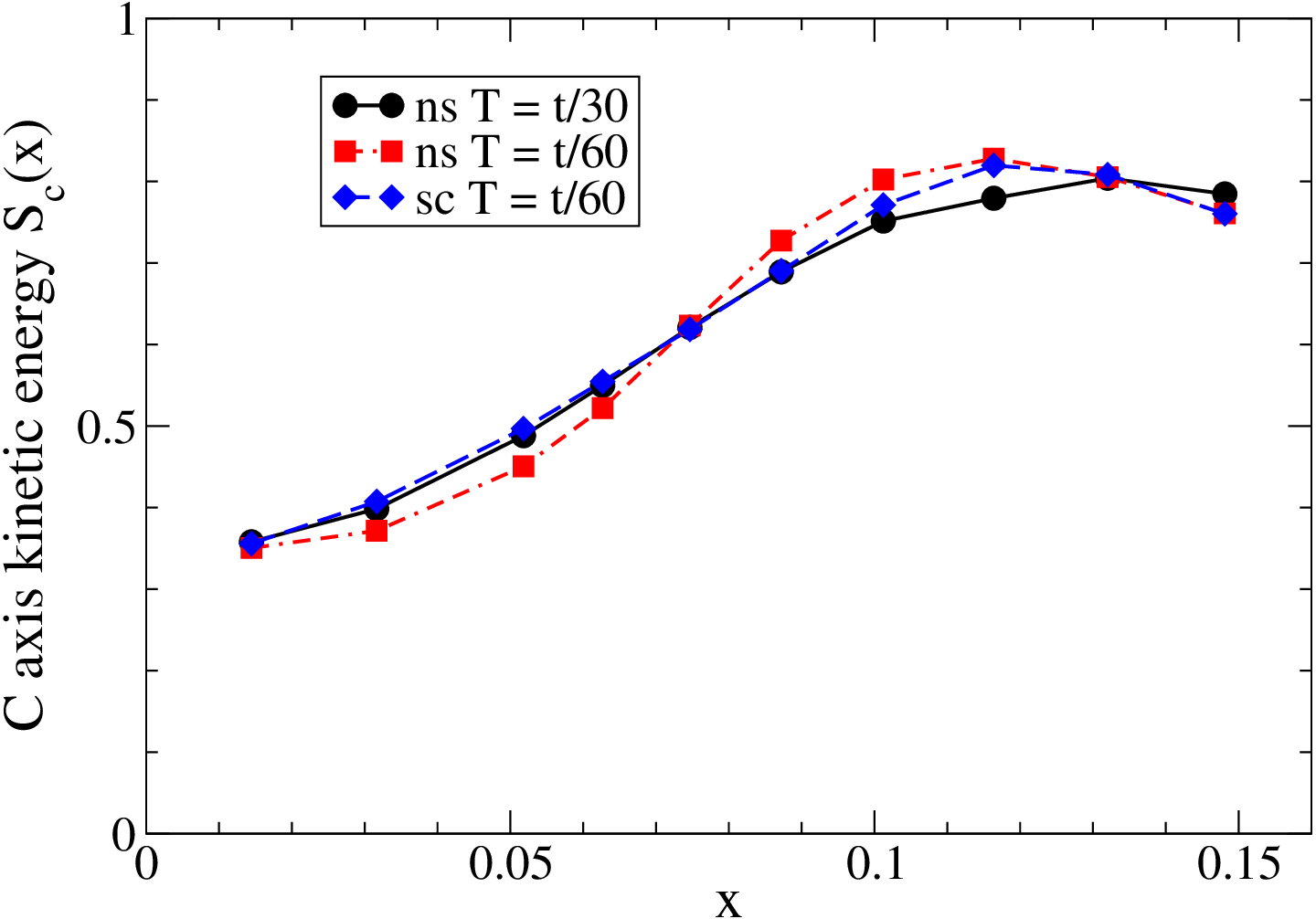}\caption{Interplane spectral weight $S_c(x)$ (optical integral obtained from Eq.~\ref{Scdef}) as a function of doping $x$, in the normal state at $T=t/30$ (circles, black online) and at $T=t/60$ in the normal state (dash-dotted lines, squares, red online) and in the superconducting state at $T/t=60$ (dashed lines, diamonds, blue online).
}
\label{fig:ekin}
\end{figure} 

Inspection of  Fig.~\ref{fig:caxis} shows that in the overdoped ($x\gtrsim 0.1$) regime, the low frequency spectral weight (integral of the conductivity over the region of the Drude peak) increases as T is decreased. On the other hand, in the underdoped regime ($x\lesssim 0.07$) the formation of the pseudogap corresponds to a decrease in the low frequency spectral weight.   To determine if the spectral weight is drawn from other frequencies we present in Fig.~\ref{fig:ekin} the kinetic energy, i.e. the integral of the optical conductivity over all frequencies. We see that for the normal state calculations, as temperature is decreased the kinetic energy indeed increases on the overdoped side  while on the underdoped side it decreases.  We attribute the effect to temperature dependent modifications of the scattering rate for the antinodal carriers. 

We now turn to the effects of superconductivity.  At $x=0.102$ where the superconductivity emerges from a more or less Fermi liquid state, the opening of the superconducting gap causes a suppression of the conductivity at low frequency (top middle panel). An increase in conductivity at a higher frequency $\sim 600~cm^{-1}$ is also evident. As the doping is decreased to the edge of the pseudogap regime ($x=0.075$) and beyond, we see that the peak in the superconducting conductivity moves rapidly to higher energy,  becoming comparable to the pseudogap energy scale. 

The existence of a superconductivity-induced peak is  interesting because the BCS coherence factors\cite{Schrieffer99} are such that the conductivity vanishes at the superconducting gap edge. Thus in  standard BCS/dirty limit calculations of the c-axis optical response of a layered superconductor\cite{Ioffe00} this increase does not occur: at $\Omega\neq0$  the superconducting state conductivity  lies below the corresponding normal-state curve. The changes therefore should be interpreted as arising from  superconductivity-induced changes in the electron scattering rate, presumably associated with the pseudogap.  Comparison of Fig.~\ref{fig:caxis} to Fig.~\ref{fig:raman} shows that  at the higher carrier concentration  the frequency scale in the peak in the superconducting state  c-axis conductivity matches  that in the Raman cross section, while at the lower dopings  the effects in the c-axis conductivity  clearly occur at the pseudogap scale, not the superconducting gap scale. 

\begin{figure}[tb]
\includegraphics[width=0.9\columnwidth]{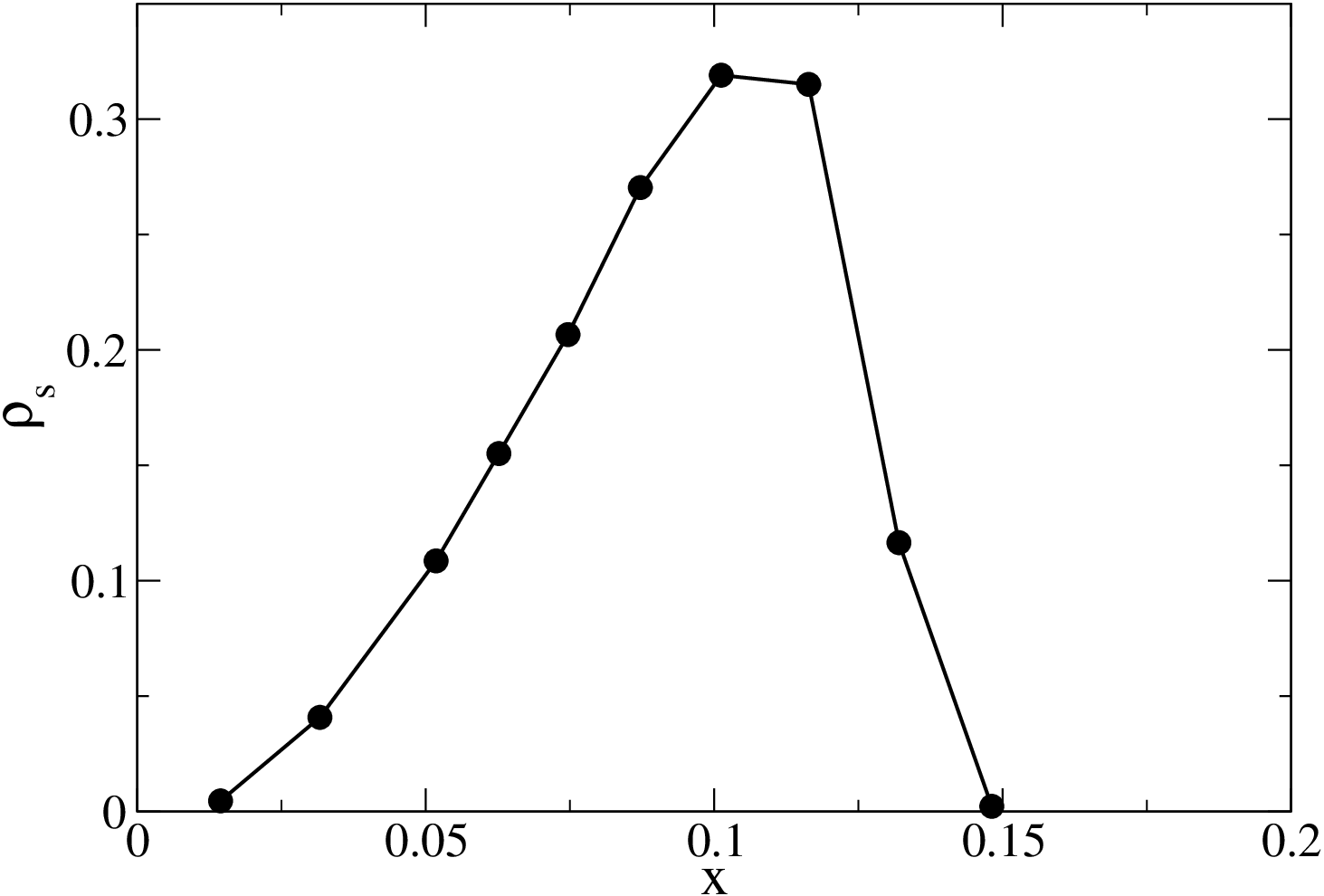}
\caption{Superfluid stiffness $\rho_s$ determined in the superconducting state at $T = t/60$ from Eq.~\ref{rhos}, as a function of doping.
}
\label{fig:rhos_n}
\end{figure}

An important characterization of the superfluid properties is the c-axis superfluid stiffness, shown for a range of dopings in Fig.~\ref{fig:rhos_n}. A clear maximum is visible.  For dopings higher than the maximum, the well defined Drude peak observed in the normal state means that conventional  clean-limit (or intermediate-scattering) BCS physics applies. The decrease in penetration depth with increasing doping in this overdoped regime occurs because our lowest accessible temperature is not far enough below the actual transition temperature, so that thermal excitations have reduced the superfluid stiffness. We believe that if the calculation could be performed at very low temperatures, the superfluid stiffness would monotonically increase with doping until the high-doping  end of the superconducting range. 

For the points $x=0.117, 0.102$ and $0.075$ the accessible temperatures are sufficiently far below $T_c$ that the results reflect the low temperature limit. It is important to note that for the value  $U=6t$ studied here, $x=0.06$ is approximately the point of maximum transition temperature and that all of these points lie outside the pseudogap regime.  The decrease in penetration depth is therefore not directly related to the pseudogap, but is due instead  to  the rapid decrease with decreasing doping  of the low frequency normal state conductivity. As discussed in Ref.~\onlinecite{Lin10} this decrease is due to the rapid increase of the low frequency scattering rate for antinodal electrons which is a precursor to the pseudogap.

Comparison of Fig.~\ref{fig:caxis} and \ref{fig:ekin} shows that for the dopings $0.12 > x >0.06$ where the low temperature limit is reached, the ``Ferrell-Glover-Tinkham'' sum rule is violated,  at about the $10\%$ level. This means that the total optical integral in the superconducting state (including the delta function contribution) is different than that in the normal state. The sign of the violation depends on doping. On the overdoped side we see that the c-axis  kinetic energy of the superconducting state is less than that in the normal state at the same temperature. On the other hand, in the underdoped state it is slightly greater.  We attribute these effects mainly to the superconductivity-induced changes in the antinodal scattering rate (on the overdoped side) and changes in the pseudogap (on the underdoped side).

\section{Discussion \label{Discussion}}
We begin the discussion by comparing our results to experimental data\cite{Cooper88,Staufer92,Chen93,Chen94,Kendziora95,Chen97,Nemetschek97,Naeini98,Opel98,Naeini99,Opel00,Hewitt02,Venturini02,Sugai00,LeTacon06,Blanc09} (for reviews see also Refs.~\onlinecite{Hackl96,Devereaux07}). As has been previously observed (see Refs.~\onlinecite{Lin10,Lin12}) the evolution of the normal-state $B_{1g}$ Raman amplitude with doping and  temperature is in good agreement with data. Overdoped materials exhibit  Raman spectra in reasonable accord with the spectra shown in the upper left panel of Fig.~\ref{fig:raman}, with a weak but visible coherent quasiparticle part which steepens as $T$ is reduced. For a doping dependent temperature  less than about $150K$  a quasiparticle peak appears, centered at $\sim 200~cm^{-1}$. As the doping is decreased, first the peak vanishes as in the upper right  panel of Fig.~\ref{fig:raman} and then the pseudogap leads to a suppression of the intensity over a wide range. See {\it e.g.} Fig.~1 of Ref.~\onlinecite{Venturini02b}.

\begin{figure}[tb]
\includegraphics[width=0.95\columnwidth]{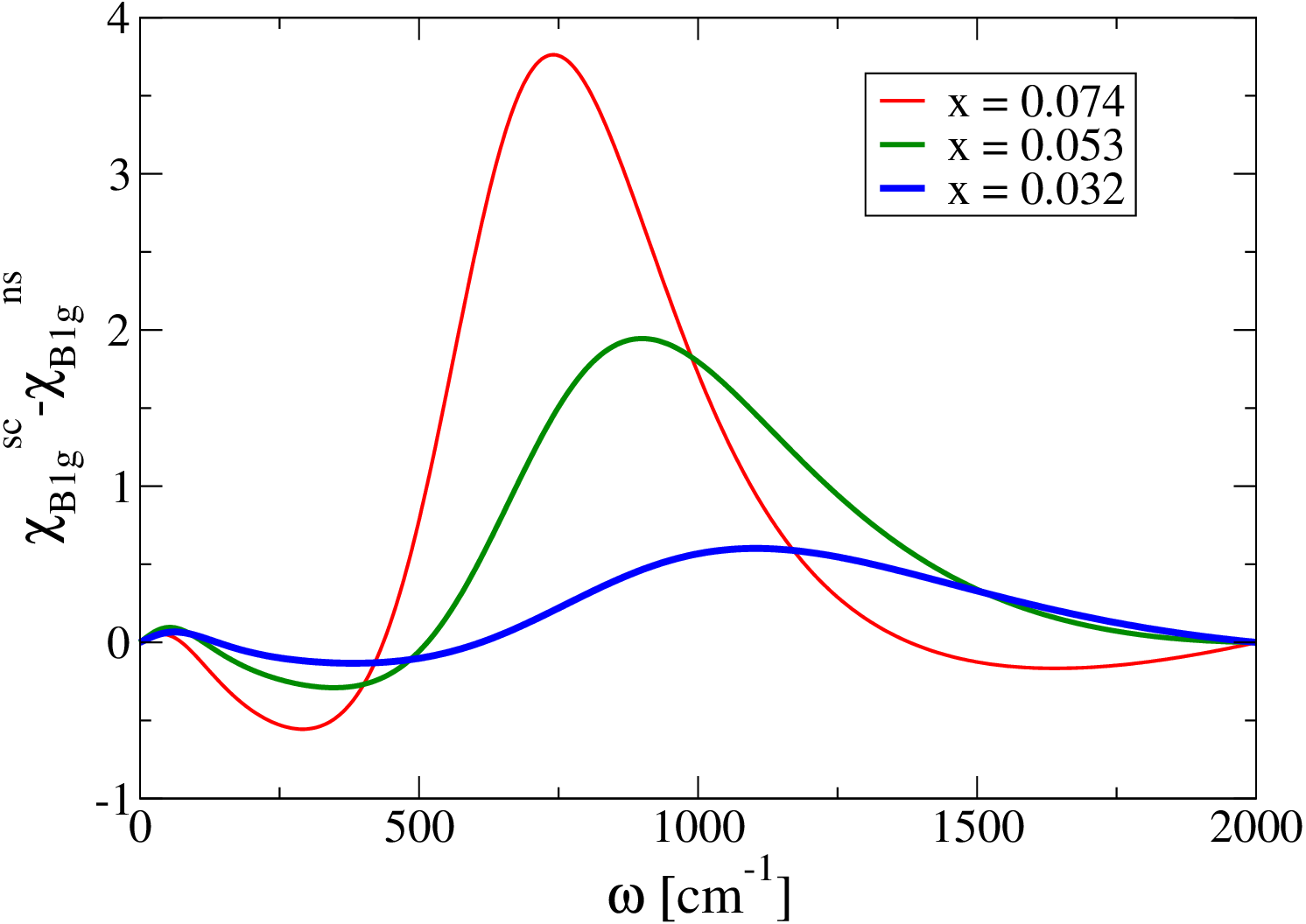}
\caption{Difference of Raman intensity computed in the superconducting state at temperature $T=t/60$ and in the normal state at temperature $T=t/30$ for carrier concentrations $x=0.075, x=0.053$, and $x=0.032$.}
\label{fig:ramansubtractnsmu-080604}
\end{figure}

The superconductivity-induced changes are similarly in reasonable qualitative agreement with the data. For overdoped materials the onset of superconductivity leads to a large amplitude peak in the Raman intensity, at a frequency which it is reasonable to interpret as twice the gap magnitude. As doping is decreased, this peak moves to higher energy, loses intensity and broadens, and at very low doping the peak  ceases to be visible. To make this more evident we plot in Fig.~\ref{fig:ramansubtractnsmu-080604}  the difference between the superconducting Raman spectra at our lowest temperature $T=t/60$, and the normal state Raman spectra at $T=t/30$.  These trends are in reasonable qualitative agreement with the results presented in Fig.~4 of Ref.~\onlinecite{Li12}. 

The theoretical curves shown in Fig.~\ref{fig:raman} reveal an additional remarkable result: at the lower dopings the size of the superconducting gap (position of the maximum in the superconducting state Raman cross section) is less than the size of the pseudogap (energy scale at which the two normal state traces merge). This phenomenon may also be observed by comparing the peak-to-peak distances in the superconducting and normal state $x=0.02$ spectral functions shown in Fig.~\ref{fig:spectra}. That the onset of superconductivity leads to a decrease in the gap scale suggests that superconductivity and the pseudogap are competing phenomena. The correspondence of the calculations to data suggest that this phenomenon also occurs in the actual materials. This idea is more extensively discussed elsewhere\cite{Gull12,Gull12b}. 

One set of  quantitative differences is that the numerical values for gaps, for the dopings delineating different regimes, and the temperature scales are somewhat different in the theory than in the experimental data. We believe this is a consequence of the relatively small value of $U=6t$ which is numerically accessible to us. The results of Ref. ~\onlinecite{Gull12} indicate that increasing $U$ will increase the doping scales and decrease the gap values.  However, Refs.~\onlinecite{Li12,Sacuto12} also report a coherence peak which remains reasonably sharp even in underdoped materials \cite{Li12}, in contrast to the broadening observed here, and also indicate that, apart from the $B_{1g}$ coherence peak, the   temperature-dependent changes are of smaller magnitude than found in the calculations reported here.  These are important issues for future investigation.

We now turn to the interplane conductivity. Here comparison with experiment\cite{Homes93,Homes95,Basov95,Timusk95,Uchida96,Homes95erratum,Schuetzmann94,Basov94,Bernhard98,Bernhard99,Dordevic04} (see also Ref.~\onlinecite{Basov05}) is complicated because in ``single layer'' compounds the coupling between the planes is so weak that it is difficult to obtain reliable data while in the $YBCO$ family of materials where the interplane conductivity is an order of magnitude larger, bilayer plasmon effects complicate the interpretation of the data. Nevertheless several important points of comparison are possible. First, the qualitative feature that the conductivity is significantly doping dependent only at frequencies below $\sim 2000 cm^{-1}$ is consistent with data (see, {\it e.g.} Fig. 2 of Ref.~\onlinecite{Dubroka10}). Second, in optimal and overdoped materials the superconductivity-induced changes correspond to a strong decrease in the absorption at frequencies below $\sim 1000cm^{-1}$ with only a weak increase at higher frequencies $>1000cm^{-1}$. As the doping is decreased the amplitude of the  changes due to superconductivity (both the low frequency decrease and the high frequency increase) rapidly become smaller.  However, an important difference is that the weak higher frequency peak does not shift as much with doping in the data as in the calculation. 

The rapid drop in superfluid stiffness as doping is decreased is consistent with observation (see, {\it e.g.}  Fig. 2a of Ref.~\onlinecite{Panagopolous00}). Our calculations reveal that the initial stages of the drop are not due to the pseudogap per se, but instead reflect the dramatic increase in low frequency zone-face scattering rate which is a precursor to the pseudgap.

\section{Conclusions \label{Conclusions}}

The high $T_c$ copper oxide superconductors exhibit both $d_{x^2-y^2}$ superconductivity, and a normal state ``pseudogap''. In some regions of phase space these phenomena coexist and manifestations in different spectroscopies of the interplay between them has been of long-standing interest in condensed matter physics.  The results of this paper, taken in conjunction with a wide range of previous work,\cite{Lichtenstein00,Maier00,Huscroft01,Parcollet04,Civelli05,Maier05,Civelli08,Macridin06,Werner098site,Gull09,Ferrero09,Ferrero09B,Ferrero10,Liebsch08,Liebsch09,Sakai09,Sakai10,Lin10,Gull10_clustercompare,Sordi10,Sordi11,Sordi12,Gull12,Sordi13} strongly suggest that  the two dimensional Hubbard model does contain the essence of the pseudogap and d-wave superconductivity phenomena observed in the cuprates.  The essential ingredient in the calculation is the electron Green function, which is affected both by the pseudogap and by superconductivity. The interplay between these two phenomena, in combination with the BCS coherence factors, leads to the somewhat different behavior in Raman and c-axis conductivities.  In the c-axis conductivity the key doping dependent changes reflect a precursor of the pseudogap, namely the rapid increase in the scattering rate for electrons near the zone face. The correspondence between the calculated and measured Raman spectra lends support to the proposal\cite{Gull12} that when superconductivity emerges from the pseudogap regime, the gap is decreased.  

The  calculations presented here are not in precise correspondence to data. Because of limits on computational resources they are performed for an interaction strength $U=6t$ and with no second neighbor hopping $t'$. The onset of the pseudogap and the maximum in $T_c$ occur at doping $x\approx 0.07$ rather than the $\approx 0.14$ observed experimentally\cite{Huefner08}.  Available evidence\cite{Lin10,Gull12} indicates that the phase boundaries move to larger $x$ as $U$ is increased, but  increasing $U$ or $t^{'}$ dramatically increases the severity of the fermion sign problem which is the crucial limiting factor in the calculation. Similarly, we studied the $N=8$ cluster dynamical mean field approximation because for   larger cluster sizes  studies at the necessary low temperatures and strong interactions are computationally too expensive. For $N=8$ (and for  $N=16$, not studied here)  the piecewise constant self energy used in the DCA dynamical mean field method produces a gap with three values: $+\Delta$ in the region of one antinode, $-\Delta$ at the other antinode and $0$ along the zone diagonal The much larger ($N=25,36$) clusters needed to resolve the details of the  momentum dependence of the gap are at present simply not accessible at the low temperatures needed for these studies.  However, results presented so far in the literature\cite{Gull10_clustercompare} make it clear that the $N=8$, $U=6$ case studied here represents  many  essential features of the model.

In summary, the physical picture emerging from these and other calculations is that the various charge-related experimental spectroscopies of the pseudogap and superconductivity may be understood in terms of the pseudogap and superconductivity effects on the electron propagator, and that these effects may reasonably be studied in terms of the two dimensional Hubbard model. The important open theoretical  issues are to extend the formalism to the computation of magnetic quantities (which requires a treatment of vertex functions) and to understand the physical origin, in the model, of the superconductivity and the pseudogap. More generally, relating these and related results to the more refined picture of the cuprates now becoming available,  in particular to the growing evidence for the important of charge and spin stripe ordering\cite{Tranquada95,Hinkov08,Taillefer10,Parker10,Ghiringhelli12} is an important open question. 

{\it Acknowledgements:} we thank A. Georges and A. Sacuto for helpful conversations. AJM was supported by NSF-DMR-1006282. A portion of this research was conducted at the Center for Nanophase Materials Sciences at Oak Ridge National Laboratory and at the National Energy Research Scientific Computing Center (DE-AC02-05CH11231), which are supported by the Office of Science of the U.S. Department of Energy. Our continuous-time quantum Monte Carlo codes are based on the ALPS\cite{ALPS20,ALPS_DMFT} libraries.

%\section{Appendices \label{Appendices}}
\appendix
\section{Numerical Methodology \label{Numerics}}
We use a momentum-space (DCA) formulation of dynamical mean field theory, written following \textcite{Maier05} in  Nambu space, thereby allowing for superconducting order\cite{Gull12}.  We restrict our solution to the paramagnetic phase, suppressing long-ranged antiferromagnetism but allowing for antiferromagnetic fluctuations (albeit coarse-grained to the momentum resolution of the cluster). We focus on clusters of size eight; a size that is a compromise between accuracy (DCA becomes exact as $N\rightarrow \infty$ but is approximate for any finite $N$) and numerical expense (away from half filling, the cost of simulations as a function of $U$, $N$, and $T^{-1}$ increases exponentially due to the fermionic sign problem). The $N=8$ cluster approximation was found in previous work\cite{Gull10_clustercompare} in the normal state to be large enough to distinguish generic $N\rightarrow\infty$  behavior from that specific to particular clusters. Work in the superconducting state\cite{Gull12} showed similar behavior but large differences to simulations on smaller clusters (both CDMFT and DCA).

The formalism requires an impurity solver formulated in Nambu space. Due to the small energy difference between the superconducting and the normal state solution\cite{Gull12b} an unbiased, numerically exact solution of the impurity model is required. Continuous-time quantum Monte Carlo methods\cite{Gull08,Gull11} are the only impurity solver methods able to access to couplings strong enough to produce a pseudogap at temperatures low enough to construct the superconducting state numerically exactly. A variant of the Continuous-time Auxiliary field (CT-AUX) impurity solver with `submatrix update' numerical techniques\cite{Gull10_submatrix} makes such simulations feasible in practice.
The solver requires a decoupling of the (repulsive) Hubbard interaction in Nambu space, which is implemented analogously to the one in normal state presented in Ref.~\onlinecite{Gull08}. The extension to Nambu space means that all matrices are twice as large as in a normal state computation at the same temperature.

We have found that the most stable procedure to obtain a converged solution is to begin at a relatively high temperature (e.g. $T = t/10$) and introduce a pairing field $\eta_1(k) = \eta_1 \phi_k$ via the replacement ${\bf G}(k,i\omega_n;\eta_1) = \left[ i\omega_n {\bf \tau}_0 + (\mu - \epsilon_k){\bf \tau}_3 + \eta_1(k){\bf \tau}_1 -{\bf \Sigma}\right]^{-1}$, with e.g. $\phi_k=\cos k_x - \cos k_y$ for $d$-wave superconductivity, and $\eta_1$ typically 0.1t. Retaining the pairing field we obtain converged solutions ${\bf G}(K,i\omega_n;\eta_1)$ first at the initial temperature, then, using the solution at the initial temperature as a seed, at the desired range of lower temperatures. We remark that the sign problem for large $\eta_1$ is less severe than at $\eta=0$, so these computations are not inordinately expensive. 

Then, at each temperature, using the converged ${\bf G}(K,i\omega_n;\eta_1)$ as a seed, we set $\eta_1=0$ in the self-consistency condition and continue iterating until convergence is reached. At selected points we check  the solution by taking the putatively converged self energy, dividing  the anomalous part by a large number (typically $20$), and verifying that under further iterations the solution converges back to the one previously found (see also supplementary material of Ref.~\onlinecite{Gull12}).

The formalism produces normal and anomalous Matsubara Green's functions $G_{A/N}(i\omega_n)$ and self-energies $\Sigma_{A/N}(i\omega_n)$, with constant relative errors of $\Sigma$: $\frac{\Delta \Sigma}{\Sigma} \sim \text{const}$ as a function of Matsubara frequency.

We then use $\Sigma_A$ and $\Sigma_N$ to evaluate the expressions for the conductivity and the stiffness (Eq.~\ref{sigmacprime} and Eq.~ \ref{rhos}) and the `bubble' term of the Raman response (Eq.~\ref{chiraman}) on the Matsubara axis. We neglect the Raman vertex corrections that have been introduced in Ref.~\onlinecite{Lin12}, based on the fact that they are (a) found to be small for small (real) frequencies, where we see the large superconducting response, and (b) at the moment too expensive to compute numerically to the accuracy required for analytic continuation.

\section{Analytical Continuation \label{Continuations}}
In the condensed matter physics context the analytical continuation problem is the inversion of a relation of the general form
\begin{equation}
M(i\Omega_n)=\int\frac{dx}{\pi}\frac{S(x)}{i\Omega_n-x}
\label{continuation1}
\end{equation}
where $M$ is a quantity measured on the imaginary frequency (or time) axis with measurement uncertainties which are relatively small, assumed to be Gaussian and characterized by a covariance matrix. $S$ is the corresponding real axis spectral function. Because the kernel $1/(i\Omega_n-x)$ has many small eigenvalues, its inversion is an ill-posed problem. With the statistical methods commonly used in condensed matter physics it is not easy to quantify the uncertainties in $S$ arising from the combination of the approximate inversion and the measurement uncertainties in $M$. We view the process as one of data fitting, which generates a spectral function that is consistent with the Matsubara data within error bars. We find that if the measurement uncertainties are sufficiently small (our relative error is typically smaller than $10^{-4}$) different implementations of the continuation process produce a reasonably robust and consistent representation of data. In general the lower frequency structures are most reliable and small differences over large frequency ranges are less robust to variations due to choices in the continuation procedure. 

In the most widely studied case, $M$ is a measurement of the electron Green function on the fermion Matsubara points $\omega_n=(2n+1)\pi T$ and $S$ is the electron spectral function, which is non-negative and normalized so that
\begin{equation}
\int_{-\infty}^\infty dx S(x)=1.
\label{Snorm}
\end{equation}
We find that our covariance matrix for $M(\Omega_n)$ is almost diagonal when measured in frequency space,\cite{Lin10} so that correlations between different bins can be neglected.

To continue the Raman scattering amplitude we observe that the Raman susceptibility also obeys a Kramers-Kronig relation
\begin{equation}
\chi(i\Omega_n)=\int \frac{dx}{\pi}\frac{\text{Im}\chi(x)}{i\Omega_n-x}
\label{chikk}
\end{equation}
but here the spectral function is odd: $\text{Im}\chi(x)=-\text{Im}\chi(-x)$.  To deal with this we  reformulate the problem as follows:
\begin{eqnarray}
\chi(i\Omega_n)&=&\int \frac{dx}{\pi}\frac{\text{Im}\chi(x)}{x}\frac{x}{i\Omega_n-x}
\\
&=&\int \frac{dx}{\pi}\frac{\text{Im}\chi(x)}{x}\left(-1+\frac{i\Omega_n}{i\Omega_n-x}\right)
\\
&=&\chi(i\Omega_n=0)+i\Omega_n\int \frac{dx}{\pi}\frac{C(x)}{i\Omega_n-x}
\label{kkS}
\end{eqnarray}
with
\begin{equation}
C(x)=\frac{\text{Im}\chi(x)}{x}.
\label{Cdef}
\end{equation}
We then invert the equation 
\begin{equation}
\frac{\chi(i\Omega_n)-\chi(i\Omega_n=0)}{i\Omega_n}=\int \frac{dx}{\pi}\frac{C(x)}{i\Omega_n-x}
\label{kkS2}
\end{equation}
using standard methods to find $C$, which is normalized according to 
\begin{equation}
\int_{-\infty}^\infty dx C(x)=\chi(i\Omega_n=0)
\label{Cnorm}
\end{equation} 
and then construct the imaginary part of the Raman response as $\chi_{Raman}=\omega C(\omega)$.

The c-axis conductivity is continued in a similar manner, defining $\chi_{\sigma_c}(\Omega_n)=\Omega_n\sigma_c^P(\Omega_n)$ and then following the steps that led to Eq.~\ref{kkS}.

We use the open source maximum entropy analytic continuation program available as part of ALPS\cite{ALPS20} for all of our continuations.

\bibliography{refs_shortened}
\end{document}